\documentstyle[epsfig,aps,prc]{revtex}

\input{psfig}

\begin{document}

\title{ Application of the information entropy to nuclei.}

\author{S.E. Massen}

\address{
Department of Theoretical Physics,
 Aristotle University of
Thessaloniki, GR 54124 Thessaloniki, Greece}

\maketitle

\begin{abstract}

Shannon's information entropies in position- and momentum- space
and their sum $S$ are calculated for various $s$-$p$ and $s$-$d$
shell nuclei using a correlated one-body density matrix depending
on the harmonic oscillator size $b_0$ and the short range
correlation parameter $y$ which originates from a Jastrow
correlation function. It is found that the information entropy sum
for a nucleus depends only on the correlation parameter $y$
through the simple relation $S= s_{0A} + s_{1A}
y^{-\lambda_{sA}}$, where $s_{0A}$, $s_{1A}$ and $\lambda_{sA}$
depend on the mass number $A$.  A similar approximate expression
is also valid for the root mean square radius of the nucleus as
function of $y$ leading to an approximate expression which
connects $S$ with the root mean square radius. Finally, we propose
a method to determine the correlation parameter from the above
property of $S$ as well as the linear dependence of $S$ on the
logarithm of the number of nucleons.
\\

 \end{abstract}

\section{INTRODUCTION}

Information theoretical methods have in recent years played an
important role in the study of quantum mechanical systems
\cite{Bialy75,Gadre84,Gadre85,Gadre87,Ohya93,%
BAB95,BAB98,Nagy96,Majer96,Panos97,Lalazi98,Massen98,%
Massen01,Massen02,Panos01b} in two cases: first in the
clarification of fundamental concepts of quantum mechanics and
second in the synthesis of probability densities in position and
momentum space. An important step was the discovery of an entropic
uncertainty relation \cite{Bialy75} which for a three-dimensional
system has the form
\begin{equation}
S= S_r + S_k \ge 3 (1+ \ln \pi) \simeq 6.434 \quad (\hbar=1),
 \label{S1}
\end{equation}
where
\begin{equation}
S_{r}=-{\int}{\rho}({\bf r}) \ln {\rho}( {\bf r})d {\bf r},
\label{Sr-1}
\end{equation}
and
\begin{equation}
S_{k}=-{\int}n( {\bf k})\ln n({\bf k})d {\bf k} \label{4}
 \label{Sk-1}
\end{equation}
are the Shannon's information entropies in position- and
momentum-space and $\rho ({\bf r})$, $n({\bf k})$ are the density
distribution (DD) and momentum distribution (MD), respectively,
normalized to unity.

Inequality  (\ref{S1}) is an information-theoretical uncertainty
relation stronger than Heisenberg's \cite{Bialy75} and does not
depend on the unit of length in measuring $\rho({\bf r})$ and
$n({\bf k})$, i.e. the sum $S=S_r+S_k$ is invariant to uniform
scaling of coordinates, while the individual entropies $S_r$ and
$S_k$ are not.  The physical meaning of $S$ is that it is a
measure of quantum-mechanical uncertainty and represents the
information content of a probability distribution, in our case of
the nuclear density and momentum distributions. Inequality
(\ref{S1}) provides a lower bound for $S$ which is attained for
Gaussian wave functions.

Information entropies were employed in the past to study
quantum-mechanical systems
\cite{Bialy75,Gadre84,Gadre85,Gadre87,Ohya93,BAB95,BAB98,Nagy96,Majer96}.
 Recently \cite{Panos97,Massen98,Massen02} we studied the
position- and momentum-space information entropies $S_r$ and
$S_k$, respectively, for the densities of various systems: the
nucleon DD of nuclei, the valence electron DD of atomic clusters
and the DD of trapped Bose alkali atoms. We found that the same
functional form $S=a+b \ln N$ for the entropy sum as function of
the number of particles $N$ holds approximately for the above
systems in agreement with Refs. \cite{Gadre84,Gadre85} for atomic
systems. In Ref. \cite{Massen01} we found a link of $S$ with the
kinetic energy of the system $T$ and a relationship of Shannon's
information entropies in position-space with an experimental
quantity i.e. the root mean square (RMS) radius of nuclei and
atomic clusters. In Ref. \cite{Lalazi98} we used another
definition of information entropy according to phase-space
considerations \cite{Ghosh84} and we derived an
information-theoretic criterion for the quality of a nuclear DD,
i.e. the larger $S$ the better the quality of nuclear model. In
Ref. \cite{Moustakidis01} the DD, the MD and the Shannon's
information entropies have been calculated for nuclei using three
different cluster expansions. The parameters of the various
expressions have been determined by least-squares fit of the
theoretical charge form factor to the experimental one. It was
found that the larger the entropy sum the smaller the value of
$\chi^2$, indicating that the maximal $S$ is a criterion of the
quality of a given nuclear model according to the maximum entropy
principle. Only two exceptions to that rule were found out of many
cases examined. Finally, in Ref. \cite{Panos01a} we considered the
single particle states of a nucleon in nuclei, a $\Lambda$ in
hypernuclei and a valence electron in atomic clusters. We proposed
a connection of $S$ with the energy $E$ of the single particle
states. Before proceeding, it is appropriate to mention that
additional applications of entropy have attracted interest in
recent years \cite{BAB95,BAB98}, but in a different spirit, in
nuclear physics problems, such as in analysis of shell model
eigenvectors. However, it is noted that in Ref. \cite{BAB98} the
authors define a correlational entropy. This is a von Neumann
entropy, which they apply in the framework of the nuclear shell
model. In our case we use the definition of information entropy
according to Shannon applied to the density distribution in
coordinate, and in momentum space of a nuclear system.

In the present work we focus on the entropy sum $S$ of a nucleus
using the analytical expressions of the DD and MD of Refs.
\cite{Massen99,Moustakidis00,Moustakidis01b}. The expressions of
those distributions have been found for $s$-$p$ and $s$-$d$ shell
nuclei using the factor cluster expansion of Clark and co-workers
\cite{Clark70} and Jastrow correlation function which introduces
short range correlations (SRC). Those expressions depend on the
harmonic oscillator (HO) parameter and the correlation parameter.
We studied the dependence of $S$ on those parameters and we found
how this quantity is connected with a fundamental quantity, the
RMS radius of the nucleus. Finally, we propose a way to determine
the correlation parameter of the model using the dependence of $S$
on that parameter as well as the linear dependence on the
logarithm of the number of nucleons. The HO parameter is
determined equating the theoretical charge RMS radius of the
nucleus with the experimental one.

The paper is organized as follows. In Sec. II, the general
definitions related to the correlated DD, MD as well as the
dependence of the entropy sum on the correlation parameter are
given. In Sec. III, we study the dependence of the RMS radius on
the correlation parameter and we give an approximate expression
connecting the information entropy sum with the RMS radius of the
nucleus. In Sec. IV, we present a method for the determination of
the correlation parameter from the information entropy sum.
Finally, in Sec. V, the summary of the present work is given.

\section{CORRELATED ONE-BODY DENSITY OF $s$-$p$ AND $s$-$d$ SHELL
NUCLEI AND THEIR ENTROPY}

A general expression for the one-body density matrix  of $N=Z$,
$s$-$p$ and $s$-$d$ shell nuclei was derived in Refs.
\cite{Moustakidis00,Moustakidis01b} using the factor cluster
expansion of Ristig, Ter Low and Clark \cite{Clark70}. That
expression depends on the HO parameter $b_0$
($b_0=(\hbar/(m\omega))^{1/2}$), the occupation probabilities of
the various states and the correlation parameter $y$ that comes
from the Jastrow type correlation function
\begin{equation}
 f(r)=1-\exp[-y r_b^2], \quad r_b=r/b_0,
 \label{Jastrow}
 \end{equation}
which introduces short range correlations. The correlation
function $f(r)$ for large values of $r$ goes to 1 and goes to 0
for $r \rightarrow 0$. It is obvious that the effect of
correlations introduced by the function $f(r)$ becomes large when
the correlation parameter $y$ becomes small and vice versa.

The diagonal part of the one-body density matrix is the DD
$\rho({\bf r})$. The Fourier transform of the DD is the form
factor
\begin{equation}
F(q)=\int \exp[i {\bf q r}] \rho({\bf r}) d {\bf r},
 \label{FF-1}
 \end{equation}
while the MD $n({\bf k})$ is given by the particular Fourier
transform of the one-body density matrix
\begin{equation}
n({\bf k})=\frac{1}{(2\pi)^3}\int \exp[i {\bf k}({\bf r} -{\bf
r'})] \rho({\bf r},{\bf r'}) d {\bf r} d {\bf r'}.
 \label{MD-1}
 \end{equation}
The expressions of $\rho({\bf r})$, $n({\bf k})$ and $F(q)$ (in
the two body approximation for the cluster expansion)  have the
forms
\begin{equation}
\rho(r)=\frac{N_0}{\pi^{3/2}b_0^3} \left[ O_1(r_b) + O_2(r_b,y)
\right],
 \label{DD-2}
\end{equation}
\begin{equation}
n(k) = \frac{N_0 b_0^3}{\pi^{3/2}} \left[ \tilde{O}_1(k_b) +
\tilde{O}_2(k_b,y) \right],
 \label{MD-2}
\end{equation}
and
\begin{equation}
F(q) = N_0 \left[ \tilde{\tilde{O}}_1(q_b) +
\tilde{\tilde{O}}_2(q_b,y) \right],
 \label{FF-2}
\end{equation}
where $k_b=kb_0$ and $q_b=qb_0$. The terms $O_1$, $\tilde{O}_1$
and $\tilde{\tilde{O}}_1$ come from the one-body term of the
cluster expansion of the one-body density matrix and the terms
$O_2$, $\tilde{O}_2$ and $\tilde{\tilde{O}}_2$ come from the
two-body term. Their expressions as well as the expression of the
normalization factor $N_0$ are given in Refs.
\cite{Massen99,Moustakidis00,Moustakidis01b}.

From the expressions of $\rho(r)$ and $n(k)$ the Shannon's
information entropies in position and  momentum space  and their
sum $S=S_r +S_k$ can be calculated through Eqs. (\ref{Sr-1}) and
(\ref{Sk-1}) for
 $\rho(r)$ and $n(k)$ normalized to 1.

For various values of the parameters $b_0$ and $y$ and for the
$N=Z$, $s$-$p$ and $s$-$d$ shell nuclei: $^4$He, $^{12}$C,
$^{16}$O, $^{24}$Mg, $^{28}$Si, $^{32}$S, $^{36}$Ar and $^{40}$Ca
we calculated $S_r$, $S_k$ and $S\equiv S_A$, treating $^{24}$Mg,
$^{28}$Si, $^{32}$S, $^{36}$Ar as $1d$ shell nuclei.

It is found that, for the above nuclei, $S_r$ and $S_k$ depend on
both parameters, $b_0$ and $y$, while their sum $S_A$ depends only
on the correlation parameter $y$. The calculated values of $S_A$
for the above mentioned nuclei versus $1/y$ are displayed by
points in Fig. 1.

It is seen that $S_A$ is an increasing function both of $1/y$ and
of the number of nucleons $A$ of the nucleus, while $S_A$ depends
almost linearly on $1/y$. For that reason we fitted the numerical
values of $S_A$ with the form
\begin{equation}
S\equiv  S_A(y) =s_{0A} + s_{1A}
\left(\frac{1}{y}\right)^{\lambda_{sA}},
 \label{SA-1}
 \end{equation}
separately for each nucleus, that is the parameters $s_{0A}$,
$s_{1A}$ and $\lambda_{sA}$ depend on the mass number $A$ of the
nucleus. The parameter $s_{0A}$ is determined from the values of
the information entropy sum in the HO case, i.e,
\[
s_{0A}=S_A(\infty),
\]
while the other two parameters are determined by least-squares fit
of the values of $S_A$ calculated from Eq. (\ref{SA-1}) to the
corresponding ones calculated from Eqs. (\ref{Sr-1}) and
(\ref{Sk-1}). The values of the parameter $s_{0A}$ and the best
fit values of the parameters $s_{1A}$ and $\lambda_{sA}$ are
displayed in Table I, while the values of $S=S_A(y)$ found from
Eq. (\ref{SA-1}) using the above values of the parameters are
displayed by lines in Fig. 1.

It is seen that the simple form of $S_A$, given by Eq.
(\ref{SA-1}), reproduces very well the numerical values of $S_A$
for all nuclei considered. Also, there is a systematic trend of
the values of the parameters $s_{0A}$, $s_{1A}$ and
$\lambda_{sA}$. The parameter $s_{0A}$ depends linearly on the
logarithm of $A$. That is expected, as $s_{0A}$ is  equal to the
information entropy sum in the HO case which depends linearly on
the  logarithm of the number of the nucleons \cite{Panos97}. The
parameter $\lambda_{sA}$ has smaller values in the closed shell
nuclei $^4$He, $^{16}$O and $^{40}$Ca than in the corresponding
neighboring open shell ones. Finally, the parameter $s_{1A}$ is
almost a monotonic increasing function of $A$ with an exception
for the nucleus $^{40}$Ca.

\section{THE DEPENDENCE OF THE INFORMATION ENTROPY ON THE RMS RADIUS}

From the analytical expression of $\rho(r)$ given by Eq.
(\ref{DD-2}) the analytical expression of the mean square radius
of the nucleus can be found. It has, in units of the HO size
parameter $b_0$, the form
\begin{equation}
r^2_b \equiv \langle r^2 \rangle /b_0^2 = R_1 + R_2(y).
 \label{MS-1}
 \end{equation}
The term $R_1$ comes from the one-body term of the one-body
density matrix and for the $N=Z$, $s$-$p$ and $s$-$d$ shell nuclei
has the form
\begin{equation}
R_1=\frac{1}{A}\left[ 6\eta_{1s} + 14 \eta_{2s} + 30\eta_{1p} +
70\eta_{1d} \right],
 \label{R1-1}
\end{equation}
where $\eta_{nl}$ are the occupation probabilities of the various
states.
 The expression of the two-body term $R_2(y)$ is given in Ref.
\cite{Massen99}. It depends on the parameter $y$ and the
occupation probabilities of the various states.

The calculated values of the RMS radius $r_b$ in units of $b_0$
versus the parameter $1/y$ for the various $s$-$p$ and $s$-$d$
shell nuclei are displayed by points in Fig. 2. As in the case of
the entropy sum $S_A$, $r_b$ is an increasing function of $1/y$
and of $A$. Although we know the analytical expression of $r_b$ we
tried to find a simpler expression for it. We found that the
simple approximate expression
\begin{equation}
r_b=r_b(y) = r_{0A} +r_{1A}
\left(\frac{1}{y}\right)^{\lambda_{rA}},
 \label{rb-1}
 \end{equation}
reproduces well the values of $r_b$ calculated from Eq.
(\ref{MS-1}). The parameter $r_{0A}$ is determined from the
relation
\[
r_{0A}=r_b(\infty) = R_1^{1/2},
\]
while the parameters $r_{1A}$ and $\lambda_{rA}$ are determined by
least-squares fit of the values of $r_b$ calculated from Eq.
(\ref{rb-1}) to the values of $r_b$ calculated from Eq.
(\ref{MS-1}).

The values of the parameter $r_{0A}$ and the best fit values of
the parameters $r_{1A}$ and $\lambda_{rA}$ for various nuclei are
displayed in Table I, while the values of $r_b$ calculated from
the fitting expression (\ref{rb-1}) using the above values of the
parameters are shown by lines in Fig. 2. It is seen that the
simple approximate expression (\ref{rb-1}) reproduces quite well
the numerical values of $r_b$. As in the case of the parameters
$s_{0A}$, $s_{1A}$ and $\lambda_{sA}$ there is a systematic trend
in the values of the parameters $r_{0A}$, $r_{1A}$ and
$\lambda_{rA}$. The parameters $r_{0A}$,  $r_{1A}$ and
$\lambda_{rA}$ are increasing functions of $A$ with exceptions in
the nuclei $^{36}$Ar and $^{40}$Ca for the parameter
$\lambda_{rA}$. Both of them are considered as $1d$ and $2s$
closed shell nuclei, respectively.

From the functional forms of $S_A(y)$
and of $r_b(y)$ given by Eqs. (\ref{SA-1}) and (\ref{rb-1}),
respectively,  the dependence of $S_A$ on the RMS radius of the
nucleus can be found. That dependence has the form
\begin{equation}
S_A(r_b) = s_{0A} + s_{1A} \left( \frac{r_b - r_{0A}}{r_{1A}}
\right)^{\lambda_{sA}/\lambda_{rA}}
 \label{Srb-1}
 \end{equation}

Thus, the information entropy sum of a nucleus has been connected
with a fundamental quantity, the RMS radius of the nucleus
measured in units of the HO size $b_0$.

\section{DETERMINATION OF THE CORRELATION PARAMETER FROM THE
INFORMATION ENTROPY }

In recent works, it has been shown that the information entropy
sum of a quantum system (electrons in atoms
\cite{Gadre84,Gadre85}, nucleons in nuclei and valence electrons
in atomic clusters \cite{Massen98} and correlated Bose atoms in a
harmonic trap \cite{Massen02}) depends approximately linearly on
the logarithm of the number of particles, given by the form
\begin{equation}
S=S(A)= a + b \ln A,
 \label{SA-2}
 \end{equation}
where $a$ and $b$ depend on the considered  system.

The question that arises is how that property can be used in
practice. A possible way is to determine $S(A)$ for two nuclei
(such that $^4$He and $^{40}$Ca) for which there are enough
experimental data and then to find $a$ and $b$ of Eq. (\ref{SA-2})
from the relations
 \begin{equation}
 a=\frac{S(4) \ln 40 -S(40) \ln 4}{\ln 40 - \ln 4},
 \label{a-1}
 \end{equation}
 \begin{equation}
 b=\frac{S(40) -S(4)}{\ln 40 - \ln 4}.
 \label{b-1}
 \end{equation}
$S(4)$ and $S(40)$ can be found calculating first the charge form
factors of $^4$He and $^{40}$Ca from the relation
\[
F_{ch}(q)= F(q) f_{cm}(q) f_p(q) f_{DF}(q),
\]
where $F(q)$ is the point form factor of the nucleus calculated
from Eq. (\ref{FF-1}), $f_{cm}(q)$ is the Tassie-Barker
\cite{Tassie58} center-of-mass correction and $f_p(q)$ and
$f_{DF}(q)$ are the correction for the finite proton size and the
Darwin-Foldy relativistic correction, respectively
\cite{Chandra76}.

The parameters $b_0$ and $y$ for $^4$He and $^{40}$Ca are
determined by least-squares fit of the theoretical $F_{ch}(q)$ to
the experimental, with the constraint that the experimental charge
RMS radius is to be reproduced. The values of the parameters $b_0$
and $y$ as well as the values of $\chi^2$ are displayed in Table
II. In the same table the values of $b_0$ and $\chi^2$ in the HO
case ($y=\infty$) are also shown. In that case $b_0$ is determined
from the above mentioned constraint. The experimental and the
theoretical $F_{ch}(q)$, calculated with and without SRC, for the
two nuclei are shown in Fig. 3.

With the values of $b_0$ and $y$ determined in the above described
way, for the two nuclei $^4$He and $^{40}$Ca, we calculated the
point $\rho(r)$ and $n(k)$ from Eqs. (\ref{DD-2}) and (\ref{MD-2})
and the Shannon's information entropies $S_r$ and $S_k$ from Eqs.
(\ref{Sr-1}) and (\ref{Sk-1}) and their sum, $S(4)$ and $S(40)$.
Substituting the values of $S(4)$ and $S(40)$ into Eqs.
(\ref{a-1}) and (\ref{b-1}), the parameters $a$ and $b$ are
determined. The calculated values
\[
a=5.4029\quad \mbox{and} \quad b=0.9360,
\]
are quite close to the values $a=5.325$ and $b=0.858$ which have
been found in Ref. \cite{Massen99} with SkIII interaction.

Rearranging Eq. (\ref{SA-1}) and replacing $S_A(y)$ by $S(A)$ from
Eq. (\ref{SA-2}), we may write
\begin{equation}
y=\left( \frac{s_{1A}}{S(A) - s_{0A}} \right)^{1/\lambda_{sA} }.
 \label{y-1}
 \end{equation}
Using  the values of the parameters $s_{0A}$, $s_{1A}$ and
$\lambda_{sA}$ given in Table I and the values of $S(A)$
calculated from Eq. (\ref{SA-2}), the  correlation parameter $y$
is determined for the other nuclei without any fit to experimental
data. The HO parameter $b_0$, can be determined now for each
nucleus from the relation
\begin{equation}
r^2_{exp} = \langle r^2 \rangle   -\frac{b_0^2}{A} + r_p^2 +
\frac{\hbar^2}{2m^2c^2}
 \label{rb-cor}
 \end{equation}
where $r^2_{exp}$ is the experimental mean square charge radius of
the nucleus  and $\langle r^2 \rangle $ is the point mean square
radius calculated from Eq. (\ref{MS-1}). The last three terms of
Eq. (\ref{rb-cor}) are the corrections due to the spurious
center-of-mass motion, the finite proton size and the Darwin-Foldy
relativistic correction, respectively.

The values of $b_0$ and $y$ for the various $s$-$p$ and $s$-$d$
shell nuclei determined in the way above described, as well as the
values of the least-squares errors, in the comparison of the
theoretical $F_{ch}(q)$ to the experimental and the entropy sum
$S$ are displayed in Table II. In the same table the values of
$b_0$, $\chi^2$ and $S$ when SRC are not included (HO case) are
also shown. From Table II we can see that there is a systematic
behavior of the parameter $y$. The values of $y$ are always
smaller (that is large correlations) in the closed shell nuclei,
$^{4}$He, $^{16}$O, $^{36}$Ar and $^{40}$Ca, than in the
neighboring open shell ones. It is mentioned that $^{36}$Ar is
treated as 1d closed shell nucleus. The above behavior indicates
that there should be a shell effect in the case of closed shell
nuclei for the values of the correlation parameter $y$. Similar
behavior was found in Ref. \cite{Moustakidis00} where the
determination of the parameters $b_0$ and $y$ were made by
individual fit of the theoretical $F_{ch}(q)$ to the experimental
one. From the same table we can see that $y$ is a monotonically
increasing function of the number of nucleons of the closed shell
nuclei.

The theoretical $F_{ch}(q)$ with and without SRC, as well as the
experimental ones for various nuclei have been plotted versus the
momentum transfer $q$ in Figs. 4 and 5. It is remarkable that
without fit to the experimental charge form factors, the present
method gives good form factors, reproducing the diffraction minima
and maxima in the correct place.  In nearly all cases, the
$\chi^2$ values found with SRC are better than the corresponding
values in the HO case. However, the assessment of the quality of
the calculated form factors should not be based solely on the
values of the least-squares errors but also on the fulfillment of
the requirement that all the diffraction minima are reproduced in
the correct place. Thus, comparing the quality of the form factors
calculated in the present method with the ones calculated with the
harmonic oscillator model, we can say that the quality of the form
factors are considerably better in the former case. All the
diffraction minima (even the third one which seems to exist in the
experimental data of $^{24}$Mg, $^{28}$Si and $^{32}$S) are
reproduced in the present method while in the harmonic oscillator
model they are not. We note also that, in the case of the nucleus
$^{36}$Ar there are not experimental data for the form factor. The
exception which appears in $^{16}$O, where the value of $\chi^2$
with SRC is worse compared with the value of $\chi^2$ without SRC,
should not be taken as a drawback of the present method. The
reason is that there exist many experimental points at low
momentum transfer where the HO model gives good form factor, while
there are a few experimental points at high momentum transfers
where the present method reproduces these points very well, as
well as all the diffraction minima and maxima. That can be seen in
Fig. 4b. If the experimental points were distributed uniformly,
then the values of $\chi^2$ calculated within the present method
would be smaller than the ones calculated within the HO model.
Thus, we should conclude that even in $^{16}$O the theoretical
$F_{ch}(q)$ calculated within the present method is better than
that calculated within the HO model.

It should be noted that in the above analysis, the open shell
nuclei, except $^{12}$C, have been treated as $1d$ shell nuclei.
However, the same analysis could be made if they were considered
as $1d$-$2s$ shell nuclei, provided that the corresponding
occupation probabilities of the various states are known. That
analysis was made, as an example, for $^{36}$Ar, assuming that the
occupation probabilities of the various states are:
$\eta_{1s}=\eta_{1p}=\eta_{2s}=1$ and $\eta_{1d}=0.8$. The values
of the parameters $b_0$ and $y$ which were found are: $b_0=1.7861\
$fm and $y=7.7658$. Thus, the value of the parameter $y$ now
becomes larger (less correlations) than that of the closed shell
nucleus $^{40}$Ca and closer to the values of the open shell
nuclei $^{24}$Mg and $^{28}$Si (see Table II).

Finally, comparing the values of the information entropy sum,
which were  calculated with and without SRC and are displayed in
Table II, it is seen that the introduction of SRC increases the
information entropy sum by $3\%$ to $5\%$ in agreement with the
simple model of SRC used in Ref. \cite{Panos97}.

 \section{SUMMARY}

In the present work a systematic study of Shannon's information
entropy sum $S$ has been made for various $N=Z$, $s$-$p$ and
$s$-$d$ shell nuclei using correlated one-body density matrix
which depends on the HO size $b_0$ and the correlation parameter
$y$.

It is found that, for all the nuclei we have examined, $S$ depends
only on $y$ through a simple two-parameter relation. A similar
approximate expression holds for the RMS radius of the nucleus
$r_b$, measured in units of $b_0$. It is found also that a simple
relation connects $S$ with $r_b$.

From the dependence of $S$ on $y$ and its linear dependence on the
logarithm of the number of nucleons of the nucleus, the
correlation parameter $y$ for a nucleus can be determined,
provided that there are enough experimental data for two
neighboring nuclei. It is mentioned that, usually, the two
parameters of the correlated one-body density matrix are
determined for each nucleus by least-squares fit of the
theoretical $F_{ch}(q)$ to the experimental. Within the present
method, those parameters are determined even in those cases where
there are not any experimental data for the charge form factor as
this has been made for the nucleus $^{36}$Ar. The only
experimental data which are used are the experimental charge RMS
radius of the nucleus, as well as the experimental charge form
factors and RMS radii of only two nuclei, those of $^4$He and
$^{40}$Ca. The procedure which is followed to determine the
parameters of the model is the following: First we determine the
parameters $a$ and $b$ of Eq. (\ref{SA-2}) from the experimental
data of $^{4}$He and $^{40}$Ca. Thus, the values of $S(A)$ for
each nucleus are known. From the value of $S(A)$ for a particular
nucleus and from Eq. (\ref{SA-1}) we determine the correlation
parameter $y$ corresponding to each nucleus. From that value of
$y$ and from Eq. (\ref{MS-1}) the parameter $b_0$ is determined so
that the theoretical RMS charge radius to be the experimental one.
It is noted also that, using the sum of the information entropies
$S_r$ and $S_k$, the short range correlation parameter has been
determined indirectly, from the density distribution, as well as
from the momentum distribution. This appears to be an interesting
feature of the present method, since that parameter is usually
determined only from the density distribution.  We would like to
mention that "experimental data" for the momentum distribution are
not directly measured but are obtained by means of y-scaling
analysis \cite{Ciofi91} and are only known for $^4$He and
$^{12}$C.

It would be interesting if the dependence of the information
entropy sum on the various parameters could be examined with more
sophisticated models, such as the density dependent interactions,
and if the present method could be applied for the determination
of a parameter of those models.

\section*{ACKNOWLEDGMENTS}

The author would like to thank Professor M.E. Grypeos and Dr. C.P.
Panos for useful comments on the manuscript.


\newpage

\begin{table}[h]
\caption{The values of the parameters $s_{0A}$, $s_{1A}$ and
$\lambda_{sA}$ of the information entropy sum $S_A$ of the
relation (10), and of the parameters $r_{0A}$, $r_{1A}$ and
$\lambda_{rA}$ of the RMS radius of the relation (13) for various
$s$-$p$ and $s$-$d$ shell nuclei.}
\begin{center}
\begin{tabular}{ l c c c c c c c c c}
 Nucleus &$s_{0A}$ &$s_{1A}$& $\lambda_{sA}$& $r_{0A}$& $r_{1A}$& $\lambda_{rA}$\\
\hline
         &        &        &        &  &  & \\
$^4$He   & 6.4342& 1.0410& 1.0064& 1.2247& 0.4740 &1.5221 \\
$^{12}$C & 7.5086& 2.1885& 1.1548& 1.4720& 1.5579 &1.8282 \\
$^{16}$O & 7.6069& 2.6464& 1.1529& 1.5$\quad\;\;$& 2.5069 &1.9549 \\
$^{24}$Mg& 8.0933& 3.7445& 1.2390& 1.6330 & 4.1480 &2.1115 \\
$^{28}$Si& 8.2096& 4.1641& 1.2548& 1.6691 & 5.2681 &2.1959 \\
$^{32}$S & 8.2901& 4.5837& 1.2659& 1.6956 & 6.2356 &2.2275 \\
$^{36}$Ar& 8.3490& 4.9578& 1.2681& 1.7159 & 6.2862 &2.1758 \\
$^{40}$Ca& 8.4347& 4.7275& 1.2208& 1.7321 & 6.5364 &2.0809 \\
         &        &        &        &  &  & \\
\end{tabular}
\end{center}
\end{table}


\begin{table}
\caption{The values of the parameters $b_0$ (in fm) and $y$, the
$\chi^2$, the RMS charge radius $\langle r_{ch}^{2}\rangle^{1/2}$
(in fm) and the information entropy sum $S$ for various $s$-$p$
and $s$-$d$ shell nuclei. The theoretical RMS charge radii are
equal to the experimental of Ref. [25].}
\begin{center}
\begin{tabular}{l l c c c c c}
 & & & & & &  \\
Nucleus &Case& $b_0$ & $y$  & $\chi^2$ & $\langle r_{ch}^{2}\rangle^{1/2}$
&  $S$\\
\hline
 & & & & & &  \\
$^{4}$He&SRC &1.2497 &3.7857  & 9.40  &1.676 &6.7068 \\
         &HO &1.3335 &$\infty$& 53.46 &1.676 &6.4342  \\
 & & & & & &  \\
$^{12}$C&SRC &1.5617 &7.1294  & 153.46 &2.471 &7.7351 \\
        &HO  &1.6108 &$\infty$& 181.31 &2.471 &7.5086 \\
 & & & & & &  \\
$^{16}$O&SRC &1.6451 &5.1782  & 417.05 &2.730 &8.0044 \\
        &HO  &1.7554 &$\infty$& 202.09 &2.730 &7.6069  \\
 & & & & & &  \\
$^{24}$Mg&SRC&1.7609 &7.8711  & 221.07 &3.075 &8.3839 \\
       &HO   &1.8222 &$\infty$& 226.41 &3.075 &8.0933 \\
 & & & & & &  \\
$^{28}$Si&SRC&1.7226 &7.8711  & 322.63 &3.086 &8.5282 \\
          &HO&1.7860 &$\infty$& 472.93 &3.086 &8.2096 \\
 & & & & & &  \\
$^{32}$S &SRC&1.7781 &7.4140  & 669.72 &3.248 &8.6531  \\
        &HO  &1.8559 &$\infty$& 850.03 &3.248 &8.2901  \\
 & & & & & &  \\
$^{36}$Ar&SRC&1.7885 &7.0790  &        &3.327 &8.7634  \\
        &HO  &1.8801 &$\infty$&        &3.327 &8.3490  \\
 & & & & & &  \\
$^{40}$Ca&SRC&1.8397 &7.1632  & 168.44  &3.479 &8.8620  \\
       &HO   &1.9526 &$\infty$& 230.60  &3.479 &8.4347 \\
\end{tabular}
\end{center}
\end{table}

\newpage

\begin{figure}
\begin{center}
\begin{tabular}{cc}
\hspace*{-1.5cm}{\epsfig{figure=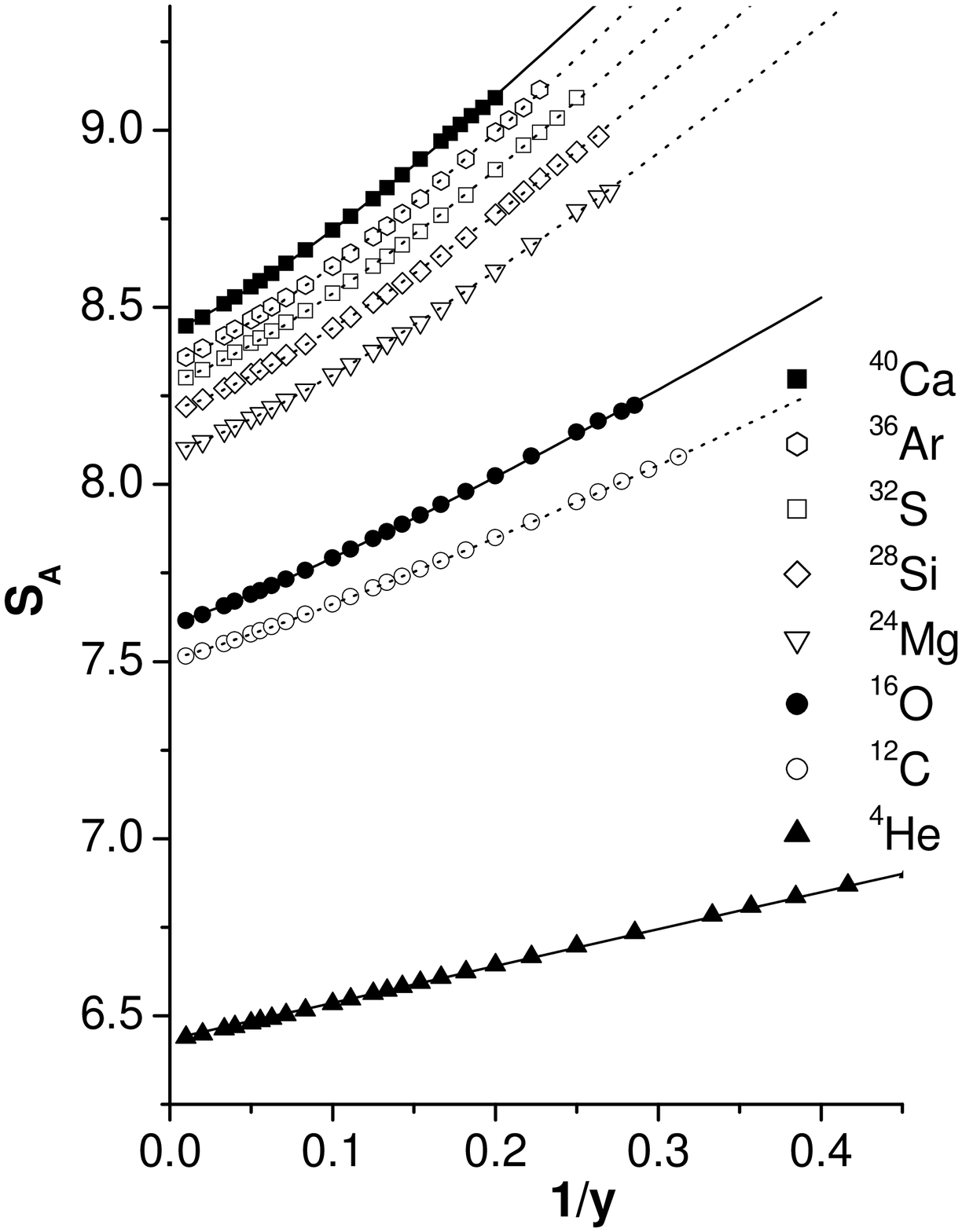,width=6.5cm} }%
\hspace*{-0.1cm}{\epsfig{figure=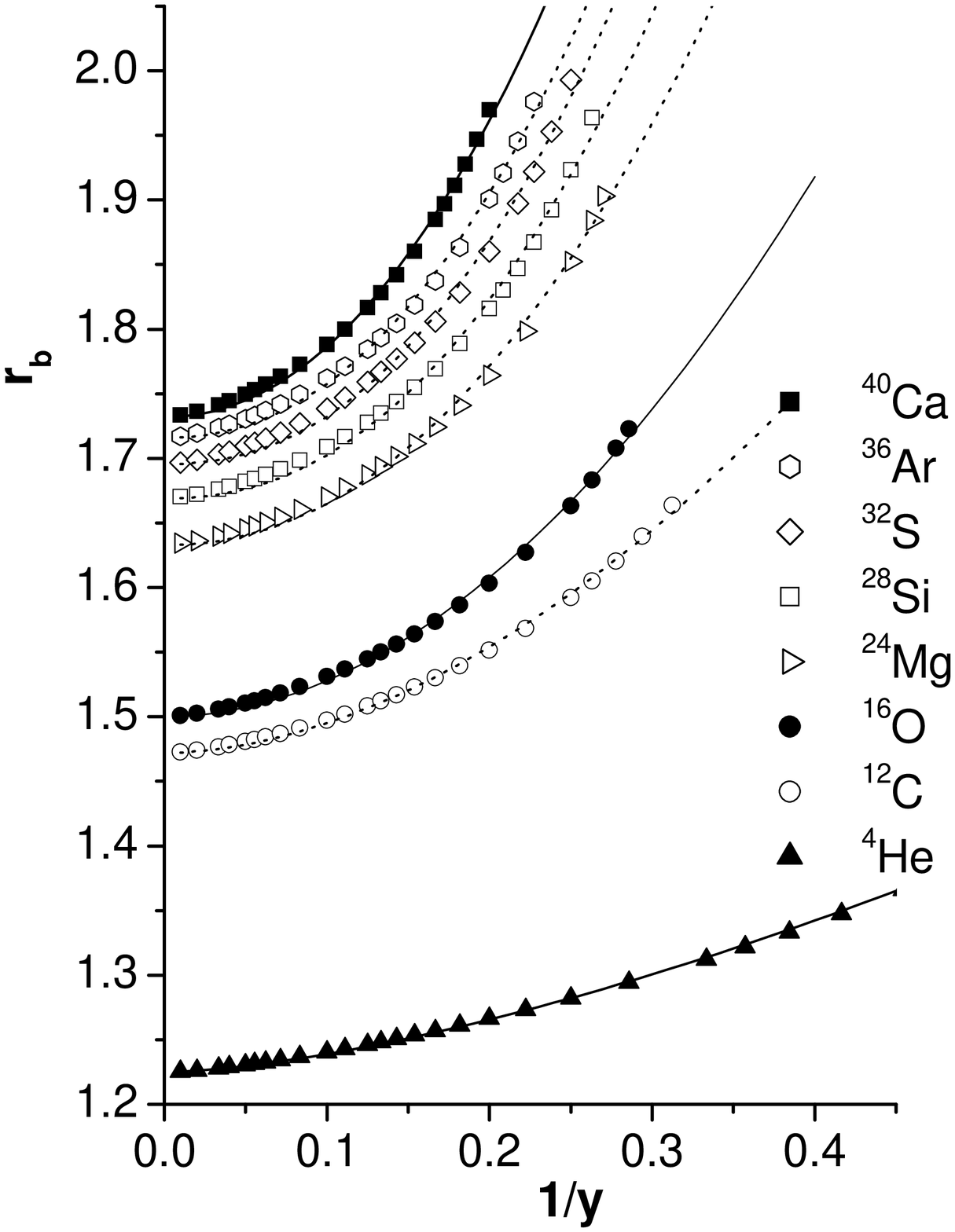,width=6.5cm} }
\end{tabular}
\end{center}
 \vspace*{-10mm}
 \hspace*{1cm}\begin{minipage}{6.cm} \caption{The
information entropy sum $S_A$ versus the correlation parameter
$1/y$ for various $s$-$p$ and $s$-$d$ shell nuclei. The points
correspond to the numerical values of $S_A(y)$ and the lines come
from the fitting expression (10).}
\end{minipage}
\hfill
 \vspace*{-2.3cm}
 \hspace*{8.cm}\begin{minipage}[b]{6.cm}
\caption{The RMS radius $r_b$, in units of the HO size $b_0$,
versus the correlation parameter $1/y$ for various $s$-$p$ and
$s$-$d$ shell nuclei. The points correspond to the values of $r_b$
calculating from Eq. (11) and the lines come from the fitting
expression (13).}
\end{minipage}
\end{figure}

\begin{figure}
\begin{center}
\begin{tabular}{cc}
\hspace*{-1.5cm}{\epsfig{figure=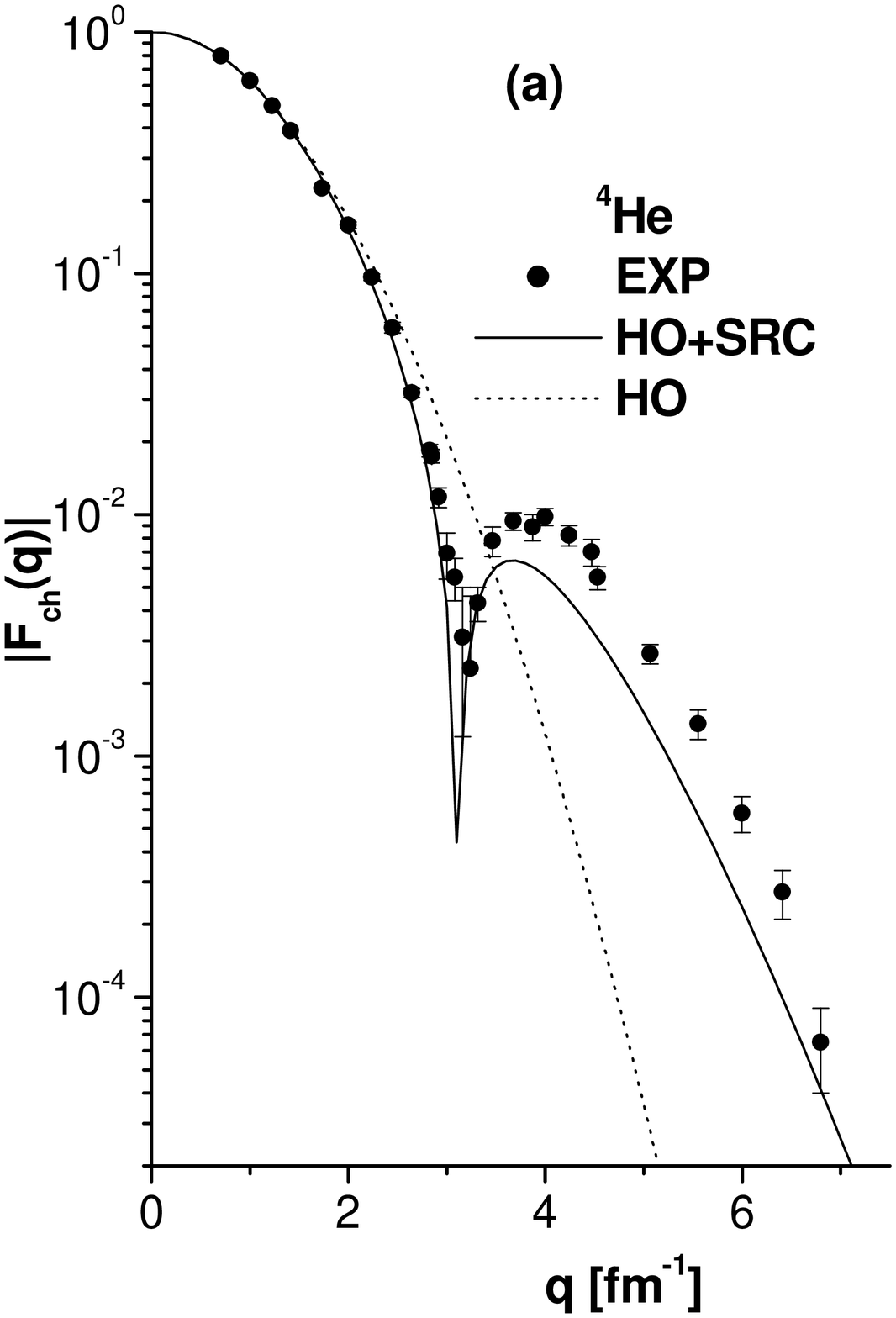,width=5.5cm} }&
\hspace*{-0.1cm}{\epsfig{figure=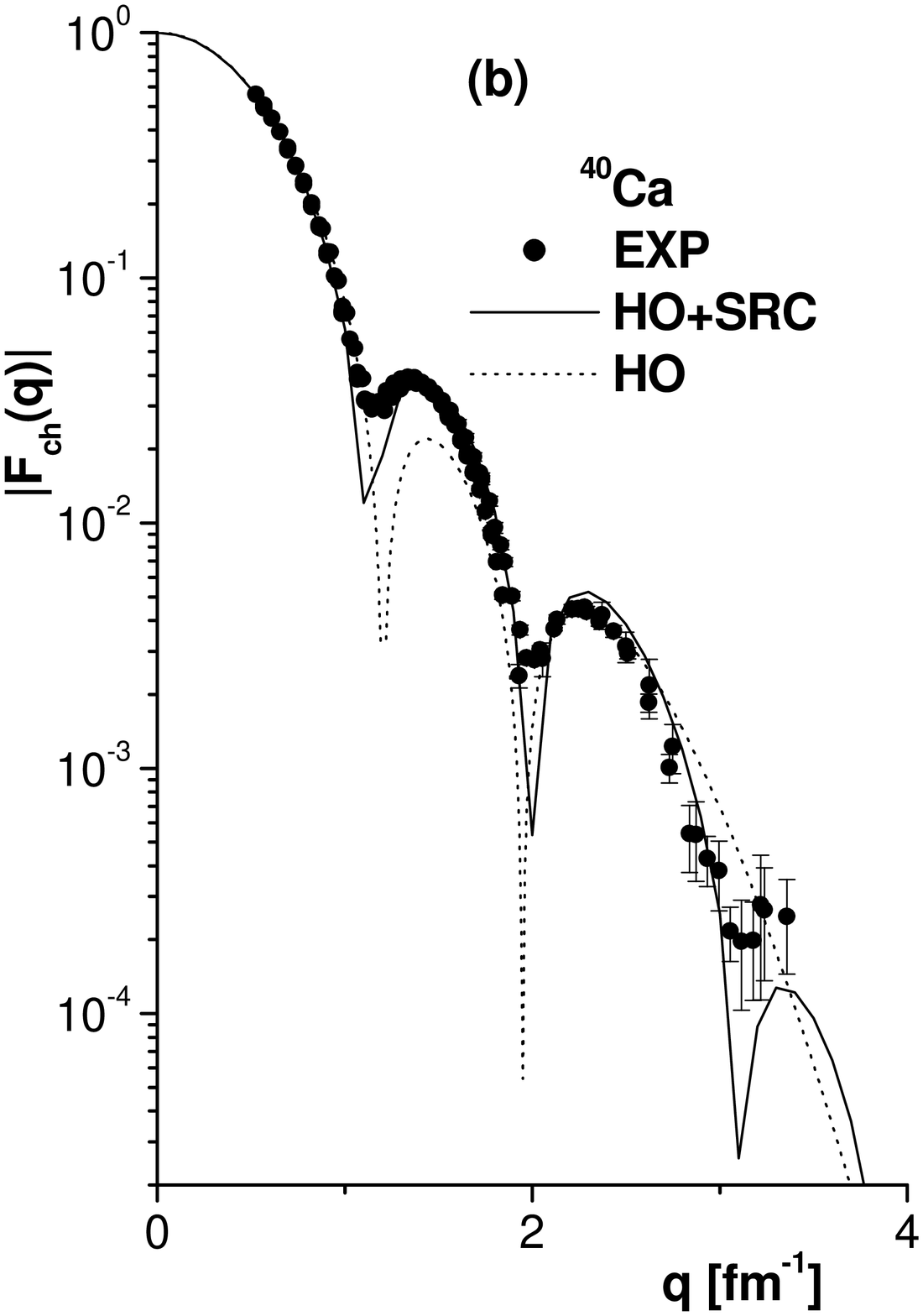,width=5.5cm} }
\end{tabular}
 \vspace*{-5mm}
\caption{The charge form factors of nuclei $^4$He (a) and
$^{40}$Ca (b). The solid lines correspond to the case when SRC are
included and the parameters $y$ and $b_0$ are determined by least
squares fit of the theoretical charge form to the experimental
with the constraint the calculated RMS charge radius is to be the
experimental one. The dot lines correspond to the HO case when
$b_0$ is determined from the experimental RMS charge radius. The
experimental points for $^4$He are from Ref. [26] and for
$^{40}$Ca from Ref. [27].}
\end{center}
\end{figure}

\begin{figure}
\begin{center}
\begin{tabular}{ccc}
{\epsfig{figure=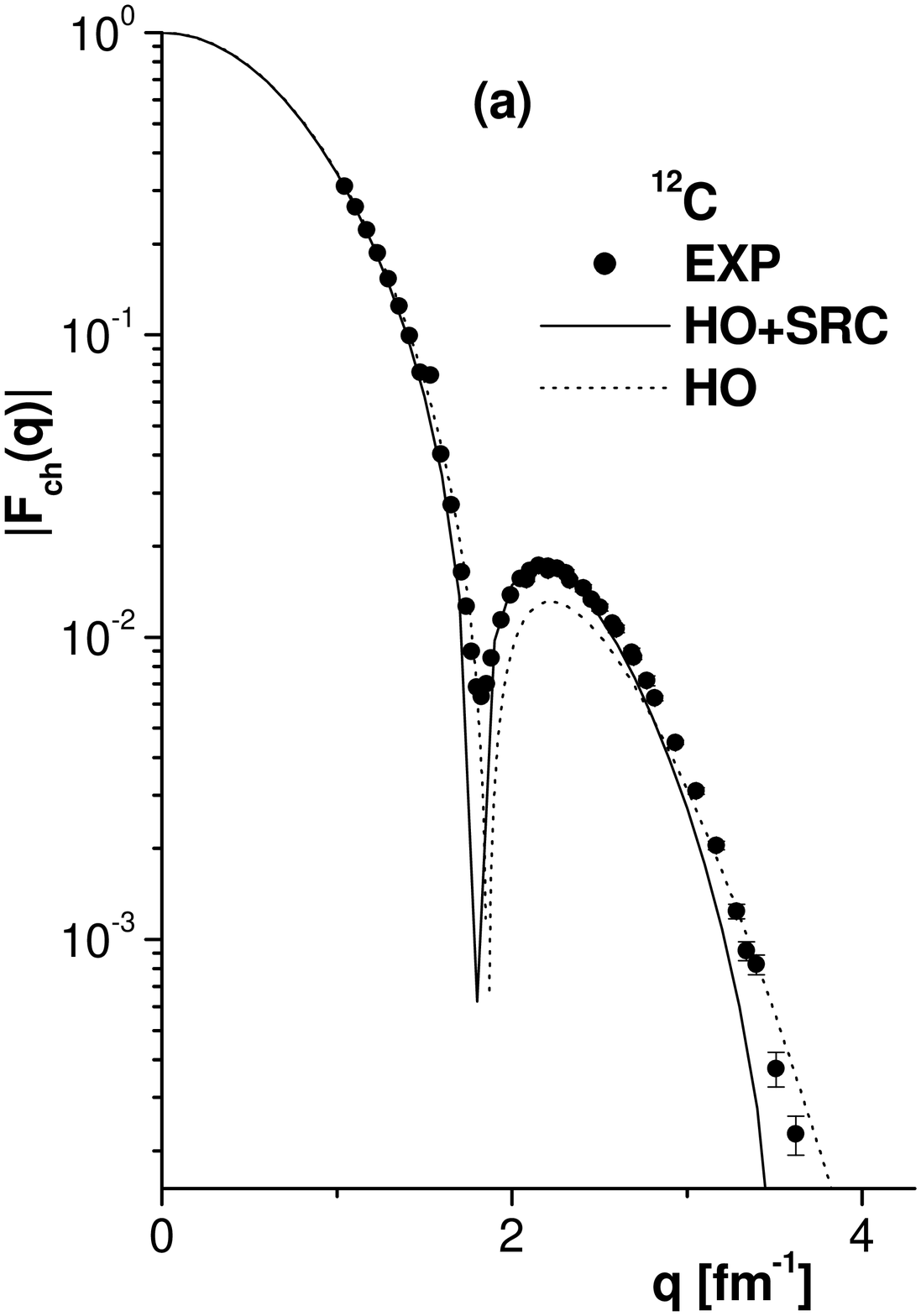,width=5.5cm} }&
{\epsfig{figure=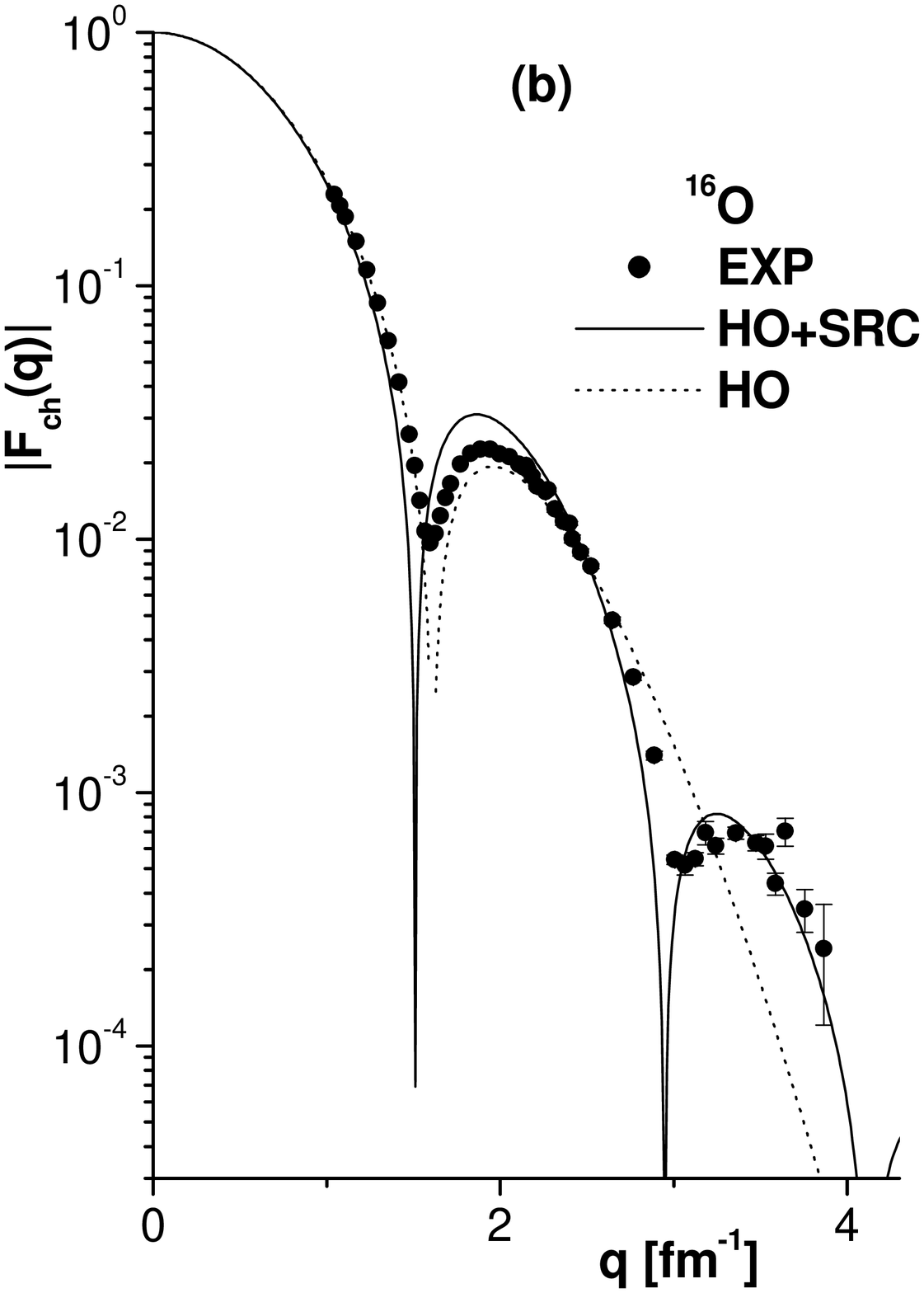,width=5.5cm} }&
{\epsfig{figure=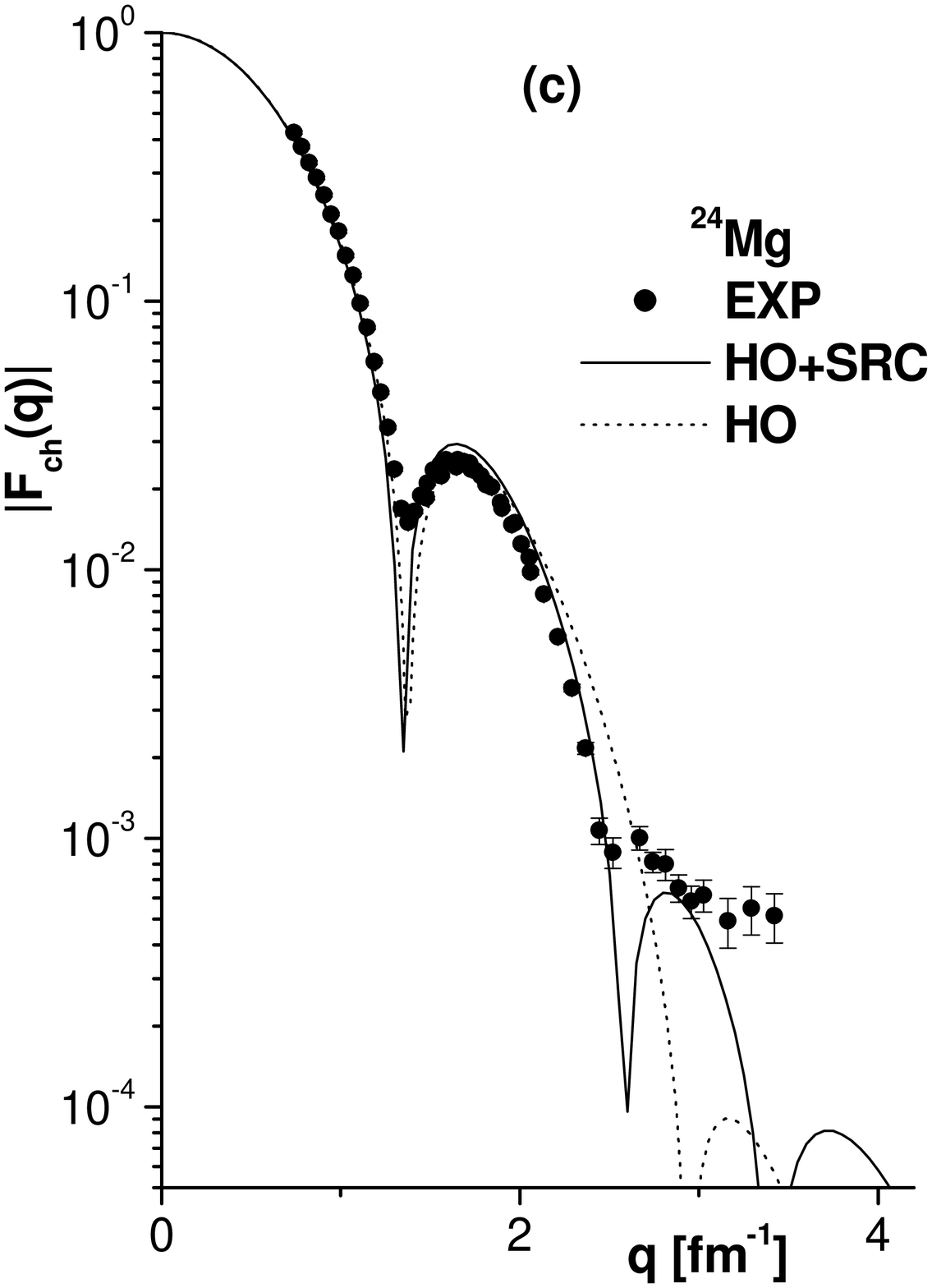,width=5.5cm} }
\end{tabular}
 \vspace*{-5mm}
\caption{The charge form factors of nuclei $^{12}$C (a), $^{16}$O
(b) and $^{24}$Mg (c). The solid lines correspond to the case when
SRC are included and the parameters $y$ and $b_0$ are determined
from Eq. (18) and the experimental RMS charge radius,
respectively. The dot lines correspond to the HO case when the
parameter $b_0$ is determined from the experimental RMS charge
radius. The experimental points for $^{12}$C and $^{16}$O are from
Ref. [28] and for $^{24}$Mg from Ref. [29].}
\end{center}
\end{figure}

\begin{figure}
\begin{center}
\begin{tabular}{ccc}
{\epsfig{figure=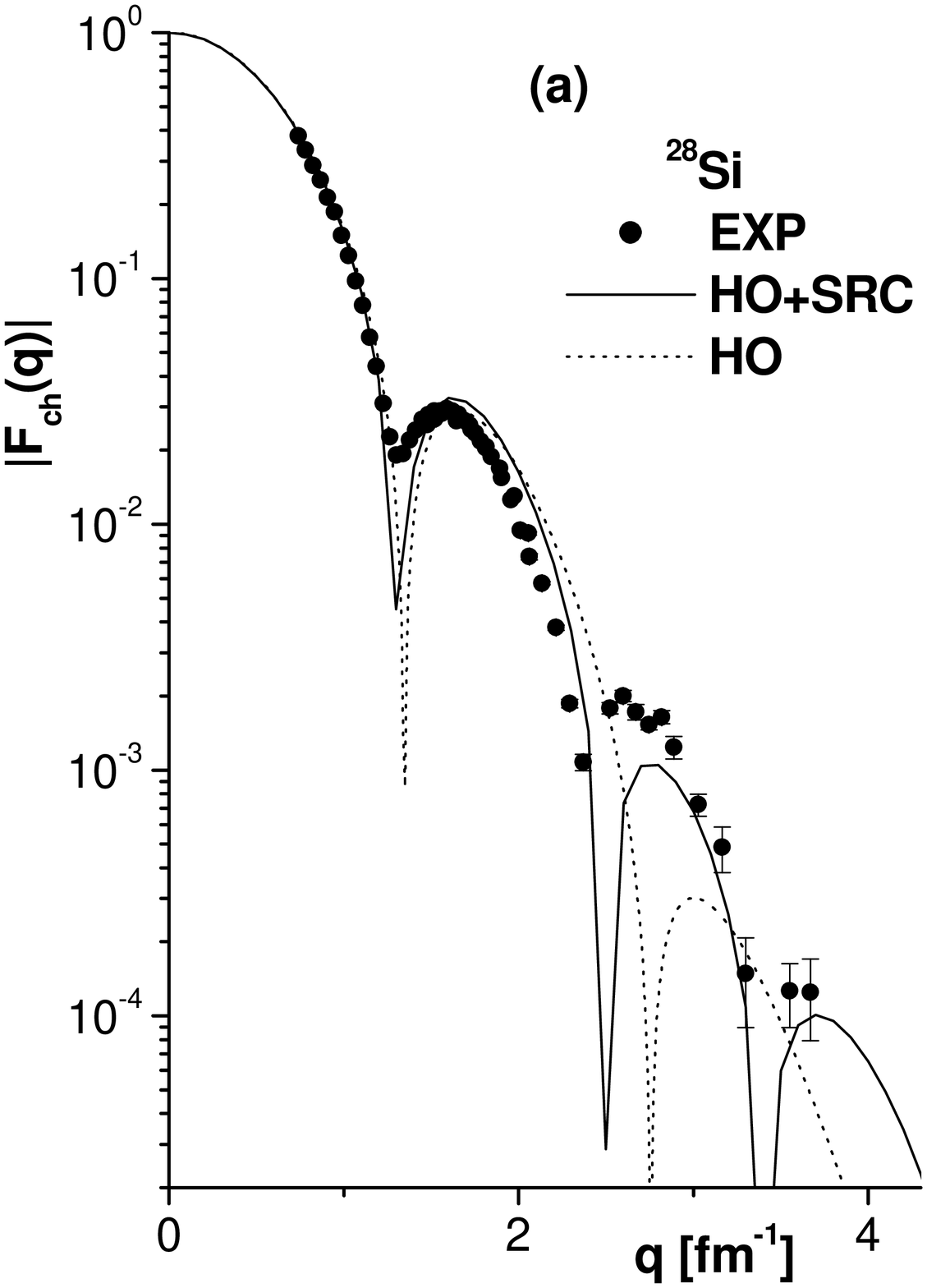,width=5.5cm} }&
{\epsfig{figure=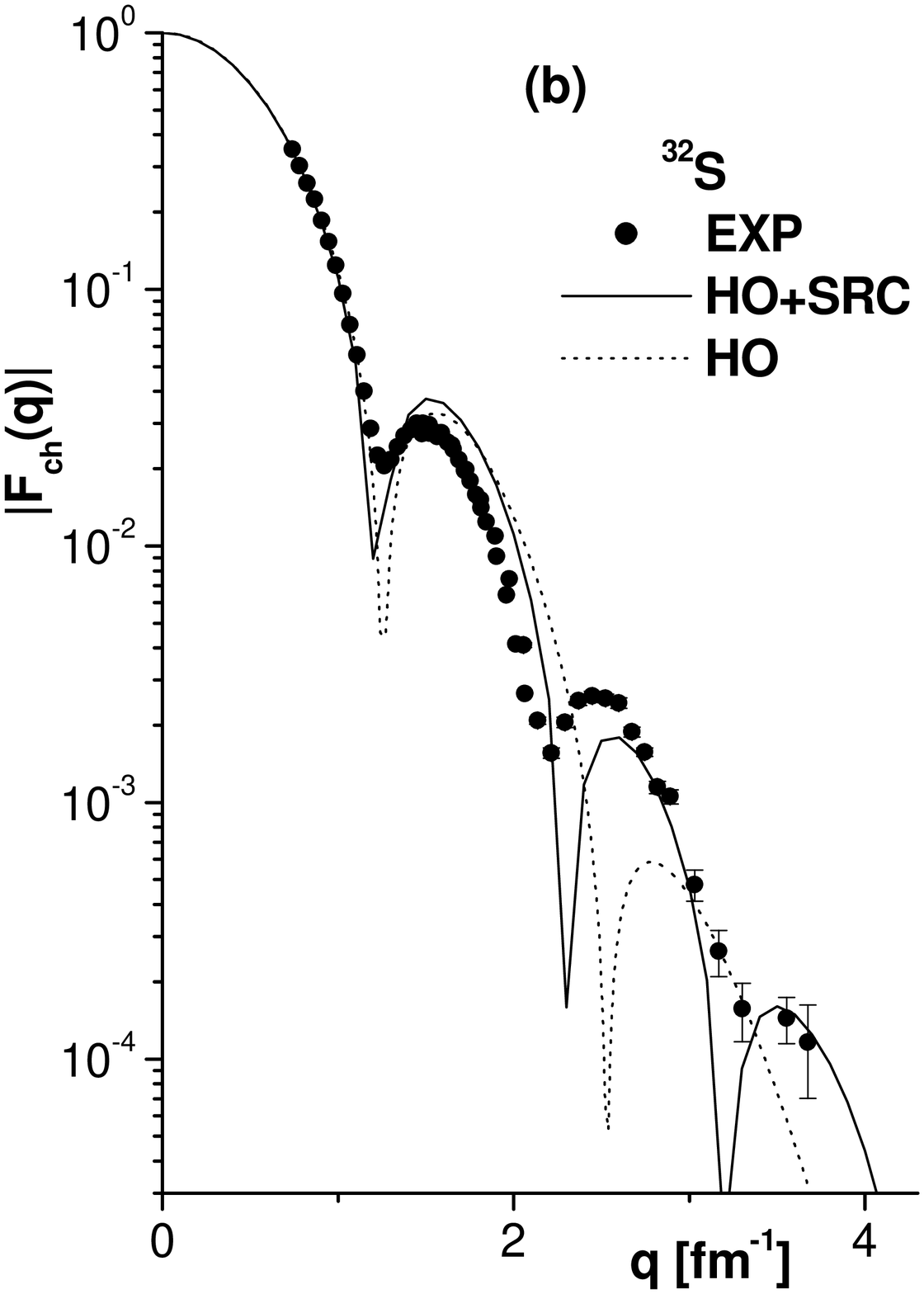,width=5.5cm} }&
{\epsfig{figure=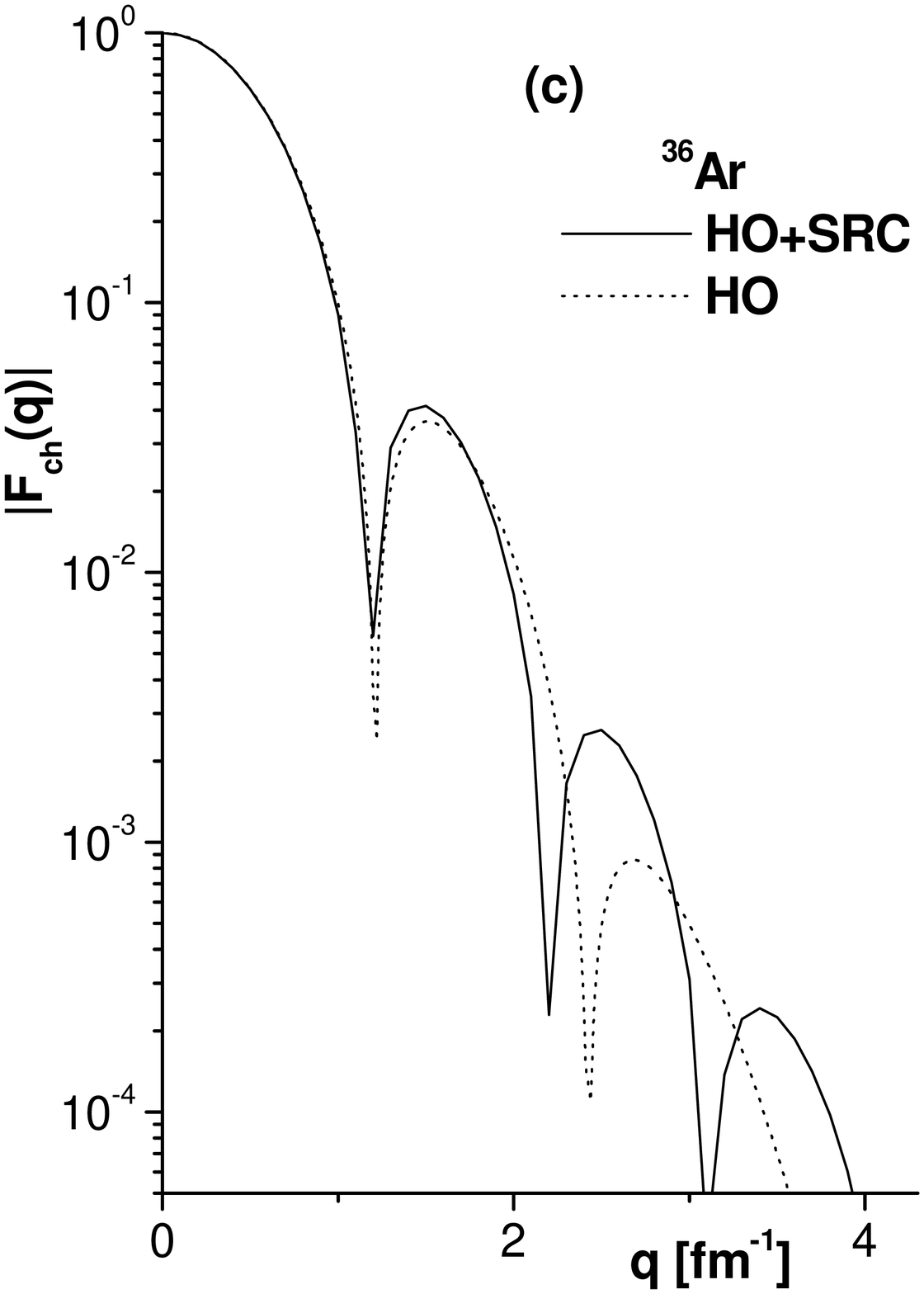,width=5.5cm} }
\end{tabular}
 \vspace*{-5mm}
\caption{The charge form factors of nuclei $^{28}$Si (a), $^{32}$S
(b) and $^{36}$Ar (c). The various cases are as in Fig. 4. The
experimental points for $^{28}$Si and $^{32}$S are from Ref.
[29].}
\end{center}
\end{figure}

\end{document}